\documentclass[amsmath,aip,rsi,preprint]{revtex4-1}
\usepackage{graphicx}

\usepackage{xcolor}

\usepackage[draft]{changes}

\makeatletter
\renewcommand\frontmatter@abstractwidth{\dimexpr\textwidth-0in\relax}
\makeatother

\begin{document}
\title{The automation of robust interatomic-force measurements}
\author{John Elie Sader}
\email{jsader@unimelb.edu.au}
\affiliation{$^1$ARC Centre of Excellence in Exciton Science, School of Mathematics and Statistics, The University of Melbourne, Victoria 3010, Australia}
\date{\today}

\begin{abstract}
\textrm{Interatomic-force measurements are regularly performed using frequency-modulation atomic force microscopy. This requires conversion of the observed shift in the resonant frequency of a force-sensing cantilever, to the actual force experienced by its tip.  Recently, Sader \textit{et al.}~Nature~Nanotechnology~\textbf{13},~1088~(2018) showed that this force conversion can be unreliable and proposed the 
\textit{inflection point test} to identify valid and robust force data. Efficient and user-friendly algorithms are required for its routine practical implementation, which currently do not exist.
Here, we (1) advance the theoretical framework of the inflection point test, (2) develop the required efficient algorithms for its complete automation, and (3) demonstrate the utility of this automation by studying two experimental datasets. The principal outcome of this report is the development of user-friendly software that integrates this automation with a standard force conversion methodology. This software provides the enabling technology  for practitioners to now seamlessly perform robust nanoscale and interatomic-force measurements.}
\end{abstract}

\raggedbottom

\maketitle

\section{Introduction}

Frequency modulation atomic force microscopy (FM-AFM) is used widely to image surfaces with atomic resolution and perform atomically-resolved force measurements.~\cite{Albrecht1991,Garcia2002,Giessibl2003} This highly sensitive technique monitors the resonant frequency of an oscillating mechanical sensor---a cantilever---as its tip is brought in proximity to a surface and experiences a force. This interaction force modifies the cantilever's stiffness and hence, in turn, the resonant frequency of the cantilever. 
Today, FM-AFM allows measurements on a multitude of samples with extreme resolution, including single organic molecules~\cite{Gross2009a} and adatoms on atomically flat surfaces,~\cite{Welker2012a} together with chemical identification of individual atoms~\cite{Sugimoto2007,Onoda2020} and operation in liquid.~\cite{Uchihashi2004}
A central requirement for all FM-AFM force measurements is the ability to convert the observed frequency shift versus distance data, into the true force experienced by the cantilever's tip. 

The frequency shift, $\Delta f(z)$, of the cantilever obeys,~\cite{Giessibl1997}
\begin{equation}
\frac{ \Delta f(z)}{f_\text{res}} = - \frac{1}{\pi a k} \int_{-1}^{1} F(z+a (1 + u)) \frac{u}{\sqrt{1-u^2}} \, du ,
\label{eq:df}
\end{equation}
where $F(z)$ is the interaction force that the tip experiences, $z$ is the tip-sample distance of closest approach, $f_\text{res}$ is the resonant frequency of the cantilever in the absence of this force, $k$ is the cantilever's dynamic spring constant and $a$ is its (fixed) tip oscillation amplitude. Equation~(\ref{eq:df}) must be inverted to recover the force, $F(z)$, from the measurement of $\Delta f$ as a function of $z$. This inversion has been studied by many authors with a range of techniques being developed. These include explicit formulas in the small~\cite{Albrecht1991} and large amplitude~\cite{Durig1999} limits, and methods that are formulated for arbitrary amplitude.~\cite{Durig2000} The latter class includes the matrix method of Giessibl~\cite{Giessibl2001} and the Sader-Jarvis method.~\cite{Sader2004}

\begin{figure}
	\includegraphics[width=0.5\columnwidth]{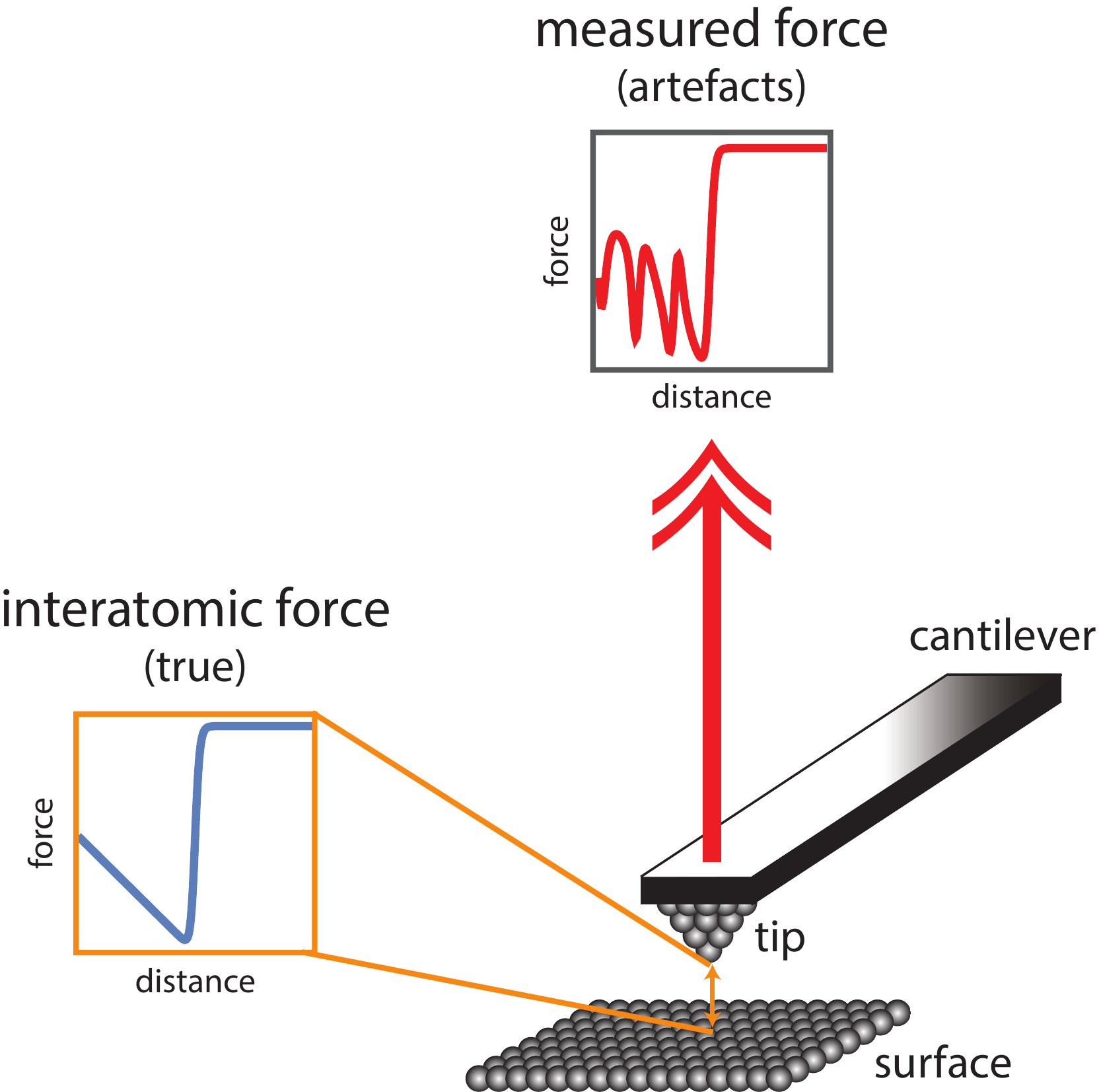}
	\caption{\label{fig:flowchart} \rm Artefacts produced by an FM-AFM force measurement.  The illustration shows a sample (true) interatomic-force law experienced by the cantilever's tip, and the measured force which contains artefacts. The presence and nature of these artefacts depends on the shape of the (true) interatomic force law and the tip oscillation amplitude.~\cite{Sader2018}}
	\label{AFMschematic}
\end{figure}

Recently,  Sader \textit{et al.}~\cite{Sader2018} found that inversion of Eq.~(\ref{eq:df}) can be ill-posed---that is, the recovered force may differ significantly from the true force and exhibit a strong dependence on (inevitable) measurement uncertainty; see example in Fig.~\ref{AFMschematic}. The reason underlying this ill-posed behavior is that the FM-AFM measurement uses a (dynamically) oscillating tip that can  blur measurement of the force, $F(z)$. That is, the tip acts as a low pass spatial filter, albeit with some subtleties.~\cite{Sader2018} If the true spatial dependence of the force varies too rapidly over the tip oscillation, details of $F(z)$ can be lost, resulting in an ill-posed inverse problem. This rapid change (jump in the force) occurs at an inflection point, where the curvature of $F(z)$ with respect to distance, $z$, changes its sign. The recovered force can then deviate from the true force.

The \textit{inflection point test} formulated in Ref.~\onlinecite{Sader2018} identifies when this ill-posedness occurs, guiding users to valid and robust force measurements. Ill-posedness produces a  zone of forbidden oscillation amplitudes, defined in Eq.~(3) of Ref.~\onlinecite{Sader2018}, that practitioners must avoid. This `forbidden zone' depends on the shape of the force law.
The inflection point test was initially demonstrated in Ref.~\onlinecite{Sader2018} and more recently in Ref.~\onlinecite{Huber2020}.

While the foundations of the inflection point test have been described previously, its routine implementation in laboratories has been restricted. This is due to the absence of user-friendly and efficient algorithms to (i) handle  multiple inflection points, and (ii) numerically evaluate higher derivatives of the recovered (discrete) force data; see the $S$-factor in Eq.~(\ref{eq:S}).

We eliminate this restriction by expanding the inflection point test's framework which is combined with new and efficient algorithms for its rapid and complete automation. In so doing, we detail how to acquire valid and robust force-distance datasets in practice and demonstrate this protocol using two discrete experimental measurements.~\cite{Huber2020} These datasets will henceforth be referred to as force-distance `curves'. Software, in the form of a Mathematica notebook, implementing this automation is provided in the Supplementary Material (along with a user guide in Section 1). This automation enables users to easily and rapidly choose an oscillation amplitude that is outside of the above-mentioned forbidden zone,  thereby ensuring a \textit{valid} force measurement. Coincidence of several valid measurements using different oscillation amplitudes indicates a \textit{robust} force measurement. The respective terms, `valid' and `robust', therefore denote measurements that are (i) insensitive to measurement uncertainty, i.e., well-posed, and (ii) independent of the oscillation amplitude chosen outside of the forbidden zone. 

\section{Theory and Automation}

In an FM-AFM force measurement, the measured $\Delta f(z)$ curve is first converted into force using a method formulated for arbitrary amplitude, e.g., Sader-Jarvis or matrix methods, discussed above.
This gives a force-distance curve whose validity is yet to be established.~\cite{Sader2018} This  force-distance curve is valid if it contains no inflection points. 
For a force-distance curve with at least one inflection point, the inflection point test provides a means for (i) assessing validity of the recovered force-distance curve, and (ii) determining the forbidden zone, and hence, the required oscillation amplitude to achieve a valid and robust force-distance curve.  The inflection point test assumes the tip-sample distance of closest approach is $z = 0$; this reference can always be set. We now summarize the inflection point test and provide a new advance that is required for its general implementation.

\subsection{Inflection point test} 

The test states that presence of an inflection point, $z = z_\textrm{inf}$, in the true force, $F(z)$, does not produce a spurious recovered force
provided the $S$-factor satisfies,~\cite{Sader2018}
\begin{equation}
	\label{eq:S}
	S(F) \equiv \frac{z_\text{inf}^2}{4} \frac{F'''(z_\text{inf})}{F'(z_\text{inf})} \gtrsim -1,
\end{equation}
where the prime symbol denotes the spatial derivative with respect to $z$. Violation of Eq.~(\ref{eq:S}) can manifest in recovered-force artefacts only in the distance range, $z \lesssim z_\textrm{inf} - 2 a$.~\cite{SJ1}

If there is just one inflection point in  $F(z)$, and Eq.~(\ref{eq:S}) holds, there is no forbidden zone and force recovery is valid for all amplitudes.
If the inequality in Eq.~(\ref{eq:S}) is violated, the recovered force-distance curve is valid when the oscillation amplitude lies outside of the forbidden zone, i.e., it satisfies,
\begin{equation}
	\label{eq:Awell}
	a \lesssim L_\text{inf} \quad \text{or} \quad  a \gtrsim \frac{z_\text{inf}}{2},
\end{equation}
where $L_\text{inf} = \sqrt{-F'(z_\text{inf}) / F'''(z_\text{inf})}$ is the length scale for the jump in $F(z)$ at the inflection point, $z_\text{inf}$. This provides a means for adjusting the oscillation amplitude, if required, to enable recovery of a valid force-distance curve.

Importantly, valid force recovery under $S(F) \lesssim -1$ occurs for all amplitudes if the force jump through the inflection point is insignificant, i.e.,
\begin{equation}
	\label{eq:jump}
	L_\text{inf} \left| F'(z_\text{inf}) \right| \ll \max \{ \left| F(z) \right| \},
\end{equation}
because ill-posedness induced at $z_\text{inf}$ exerts a negligible effect overall in this case. The necessary and sufficient condition in Eq.~(\ref{eq:jump}) has not been reported  previously.~\cite{Sader2018,Huber2020}

When the amplitude cannot be adjusted according to Eq.~(\ref{eq:Awell}), the recovered force is guaranteed to be valid for $z \gtrsim z_\text{inf}^\text{max} - 2 a$ only, where $z_\text{inf}^\text{max}$ is the position of the inflection point at greatest distance that does not satisfy Eq.~(\ref{eq:jump});  derived from the artefact distance range specified immediately after  Eq.~(\ref{eq:S}).
This property enables derivatives of $F(z)$ to be computed from the recovered force and implementation of the inflection point test.

\begin{figure}[h]
	\includegraphics[width=0.78\columnwidth]{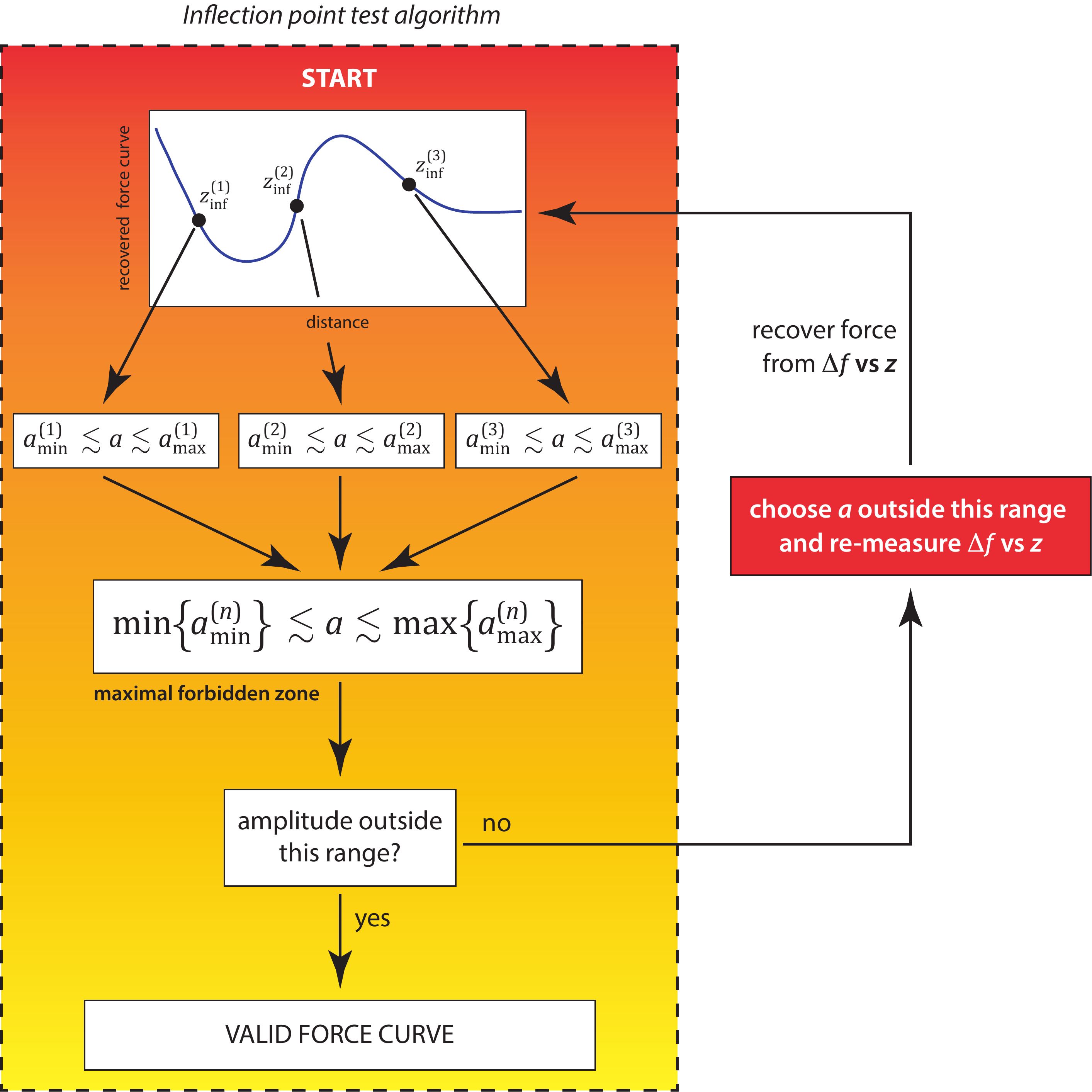}
	\caption{\label{fig:flowchart} \rm Schematic of the infection point test algorithm. The inflection point test is applied to all inflection points of a recovered force-distance curve, from which a maximal forbidden zone is determined. This maximal forbidden zone is used to select a new measurement amplitude and the FM-AFM measurement taken again to obtain a new recovered force-distance curve. The  procedure is repeated until the chosen amplitude lies outside of the maximal forbidden zone. This produces a valid force curve from which the final forbidden zone is determined. Coincidence of several valid force curves using different amplitudes identifies a robust force curve. \textit{Dashed box}: Inflection point test algorithm---software is provided in the Supplementary Material, incorporating force recovery using the Sader-Jarvis method.~\cite{Sader2004}}
\end{figure}

When the recovered force-distance curve contains more than one inflection point, the inflection point test must be applied to each inflection point and their forbidden zones determined. In principle, this is to be implemented sequentially as in Ref.~\onlinecite{Huber2020}, by (i) first applying the inflection point test at $z = z_\text{inf}^\text{max}$, (ii) adjusting the amplitude according to Eq.~(\ref{eq:Awell}) and performing a new measurement, if needed, and (iii) repeating this procedure systematically for all other inflection points (one-by-one in descending order). This can represent a significant undertaking, which we now address through the formulation of a new algorithm.

\subsection{Efficient algorithm for inflection point test} 

A more efficient algorithm is to (i) apply the inflection point test simultaneously to all inflection points of a recovered force-distance curve, i.e., at $z = z_\text{inf}^{(n)}$, where $n = 1, ... , \, N$, and $N$ is the total number of inflection points, (ii) determine their individual $S$-factors, i.e., $S_n(F)$, and thus (iii) compute their individual forbidden zones,
\begin{equation}
	\label{eq:IFTrange1}
	a_\text{min}^{(n)} \lesssim a \lesssim a_\text{max}^{(n)},
\end{equation}
where $a_\text{min}^{(n)} = z_\text{inf}^{(n)}/(2 \sqrt{-S_n(F)}\, )$ and $a_\text{max} =  z_\text{inf}^{(n)}/2$, which exist only if $S_n(F) \lesssim -1$. 

The required \textit{maximal} amplitude range, i.e., the `maximal forbidden zone', for which artefacts in the recovered force can occur is then given by that spanned by all individual zones in Eq.~(\ref{eq:IFTrange1}), i.e.,
\begin{equation}
	\label{eq:IFTrangeFINAL1}
	a_\text{lower} \lesssim a \lesssim a_\text{upper},
\end{equation}
with $a_\text{lower} = \min \{ a_\text{min}^{(n)}  \}$ and $a_\text{upper} = \max \{ a_\text{max}^{(n)}  \}$.

If the measurement amplitude satisfies Eq.~(\ref{eq:IFTrangeFINAL1}), it is adjusted to lie outside of this range and the FM-AFM measurement is redone. Then, Eq.~(\ref{eq:IFTrangeFINAL1}) is applied to the resulting (new) recovered force, and the entire procedure systematically repeated until the chosen measurement amplitude sits outside of Eq.~(\ref{eq:IFTrangeFINAL1}). This ensures that a valid force-distance curve is acquired, from which the (final) complete forbidden zone is determined.

While the maximal forbidden zone, Eq.~(\ref{eq:IFTrangeFINAL1}), may encompass amplitudes that also give well-posed behavior, this algorithm can strongly reduce the number of required measurements relative to the alternate algorithm discussed above.~\cite{Huber2020} It will achieve the same goal: a \textit{valid} (i.e., well-posed) force-distance curve.

This new algorithm, which is shown schematically in Fig.~\ref{fig:flowchart}, together with the requirement that the recovered valid force-distance curve is independent of the chosen amplitude, constitutes a robust measurement protocol. Force measurements that use only a single amplitude with no check of ill-posed behavior, as was accepted practice previously, should no longer be used and can yield incorrect force measurements.~\cite{Sader2018}

\subsection{Numerical derivatives of $\boldsymbol{F(z)}$}

A central requirement of the inflection point test is numerical evaluation of $F'(z_\text{inf})$ and $F'''(z_\text{inf})$. Since $F(z)$ arises from  experiment, and is inherently discrete/noisy in nature, care must be taken to properly evaluate these derivatives (and estimate their uncertainties). This was performed in Ref.~\onlinecite{Huber2020} using splines that require user input/judgement and no error estimate was provided.

Here, we  use the property that these derivatives occur at the inflection points, $z = z_\text{inf}$, and employ a local Taylor series representation for $F(z)$,
\begin{equation}
	\label{eq:Taylor}
	F(z) \approx F_0 + F_1 \left( z - z_0 \right) + \frac{F_3}{6} \left( z - z_0 \right)^3,
\end{equation}
where $F_0$, $F_1$, $F_3$, and $z_0$ are variables to be determined; each variable specifies a unique feature of the force-distance curve (see Eq.~(\ref{eq:Taylorres})), facilitating fits to the discrete data. Importantly, Eq.~(\ref{eq:Taylor}) is chosen to imbed the inflection point property, $F''(z_\text{inf})=0$. Equation~(\ref{eq:Taylor}) is fit directly to the discrete (and noisy) force-distance curve in the neighborhood of each individual inflection point, using a nonlinear least squares procedure. This  regularizes the derivative calculation and yields the following estimates,
\begin{equation}
	\label{eq:Taylorres}
	z_\text{inf} \approx z_0, \quad F'(z_\text{inf}) \approx F_1, \quad  F'''(z_\text{inf}) \approx F_3.
\end{equation}
This fit procedure is most readily implemented in standard software packages, which automatically provide  associated uncertainties for the estimates in Eq.~(\ref{eq:Taylorres}); see Section 2 of Supplementary Material for examples.

\subsection{Software}

A Mathematica notebook that completely automates implementation of the above-described inflection point test algorithms, together with the Sader-Jarvis method for force recovery, is  provided in the Supplementary Material. Users simply input their measured frequency-distance data, cantilever tip amplitude, stiffness and resonant frequency, from which a recovered force and its validity are reported (figure and CSV file) along with guidance on the appropriate choice of amplitude for any additional measurement. No other user input is required.

\section{Demonstration of automation}\label{demonstration}

\begin{figure*}
	\includegraphics[width=\columnwidth]{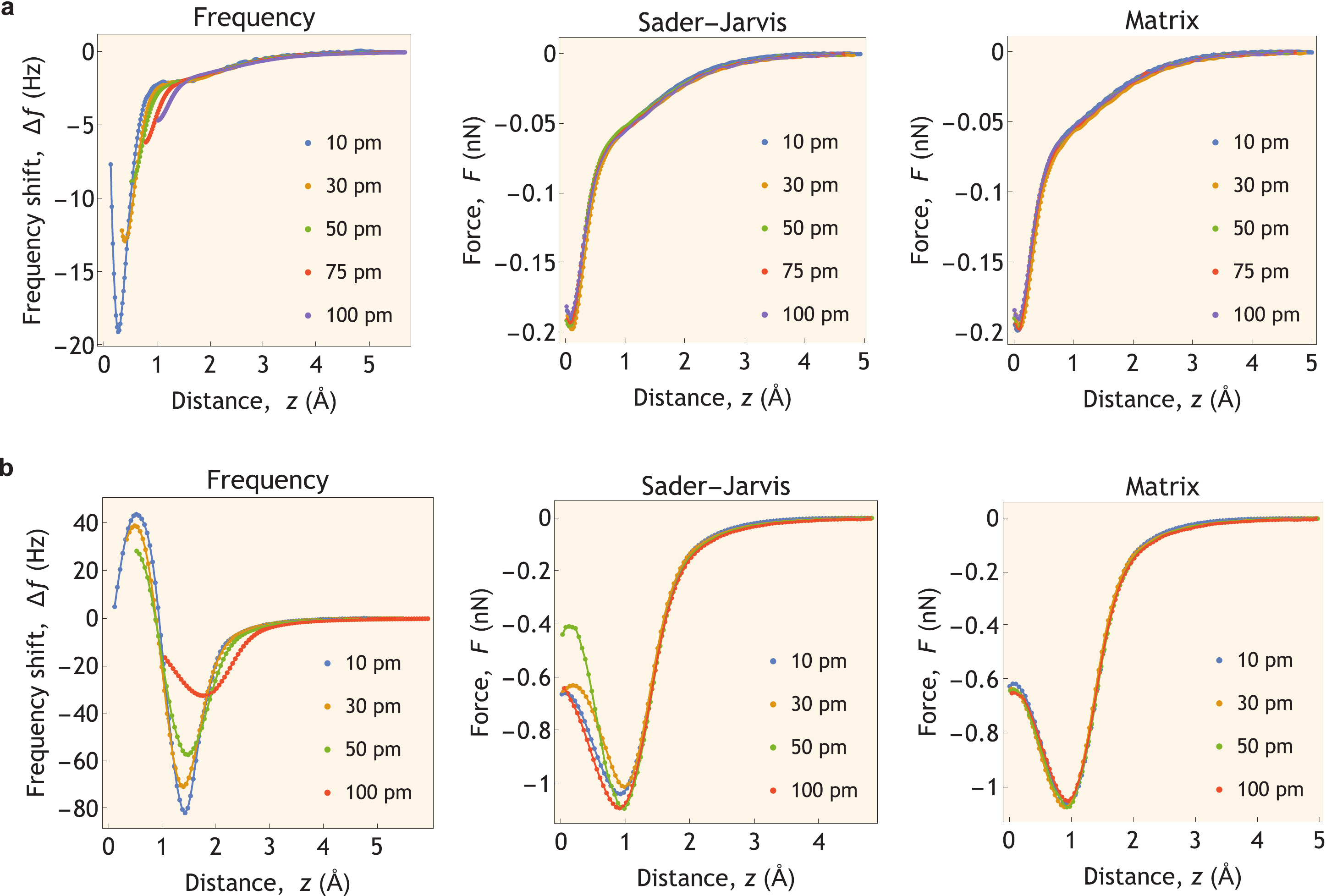}
	\caption{ \rm Sample FM-AFM force measurements from Ref.~\onlinecite{Huber2020}, used to demonstrate the inflection point test automation. These figures are redrawn with the Sader-Jarvis and matrix method data separated to highlight their differences and enable discussion in this report. \textbf{a}, CO molecule, \textbf{b}, Fe trimer, functionalized tips both over a Cu adatom on a Cu(111) surface. The primary observable in the measurement is the frequency shift, which is inverted to give the force using the Sader-Jarvis and matrix methods, as indicated.}
	\label{fig:expHuber}
\end{figure*}

We now demonstrate this inflection point test algorithm and its automation on the two experimental (discrete) data sets reported in Ref.~\onlinecite{Huber2020}: $\Delta f(z)$ spectra over the center of (i) a carbon monoxide (CO) molecule and (ii) a single Fe trimer on Cu(111), measured using a metal tip. Experiments in Ref.~\onlinecite{Huber2020} were performed using a custom-built, combined scanning tunneling and atomic force microscope operating at a temperature of 5.9\,\,K.

\subsection{CO/Cu(111) dataset}

First consider the  measurement over the center of a CO molecule adsorbed on Cu(111).
To demonstrate utility of the automated inflection point test, we use the measured short-range~\cite{Lantz2001} $\Delta f(z)$ curves in Fig.~\ref{fig:expHuber}(a)(left), for five different oscillation amplitudes. Note that  the Sader-Jarvis method code in the provided software (Supplementary Material) and separate in-house code for the matrix method give commensurate recovered force curves to those reported in Ref.~\onlinecite{Huber2020}.~\cite{SJ4} We use the latter in-house code here for consistency, which is achieved by bypassing the Sader-Jarvis code in the provided software. The inflection point test will give results independent of the chosen force recovery method, if the method is formulated for arbitrary amplitude and the recovered force curve is valid;~\cite{Sader2018} this is demonstrated below.

\begin{figure*}
	\includegraphics[width=0.75\columnwidth]{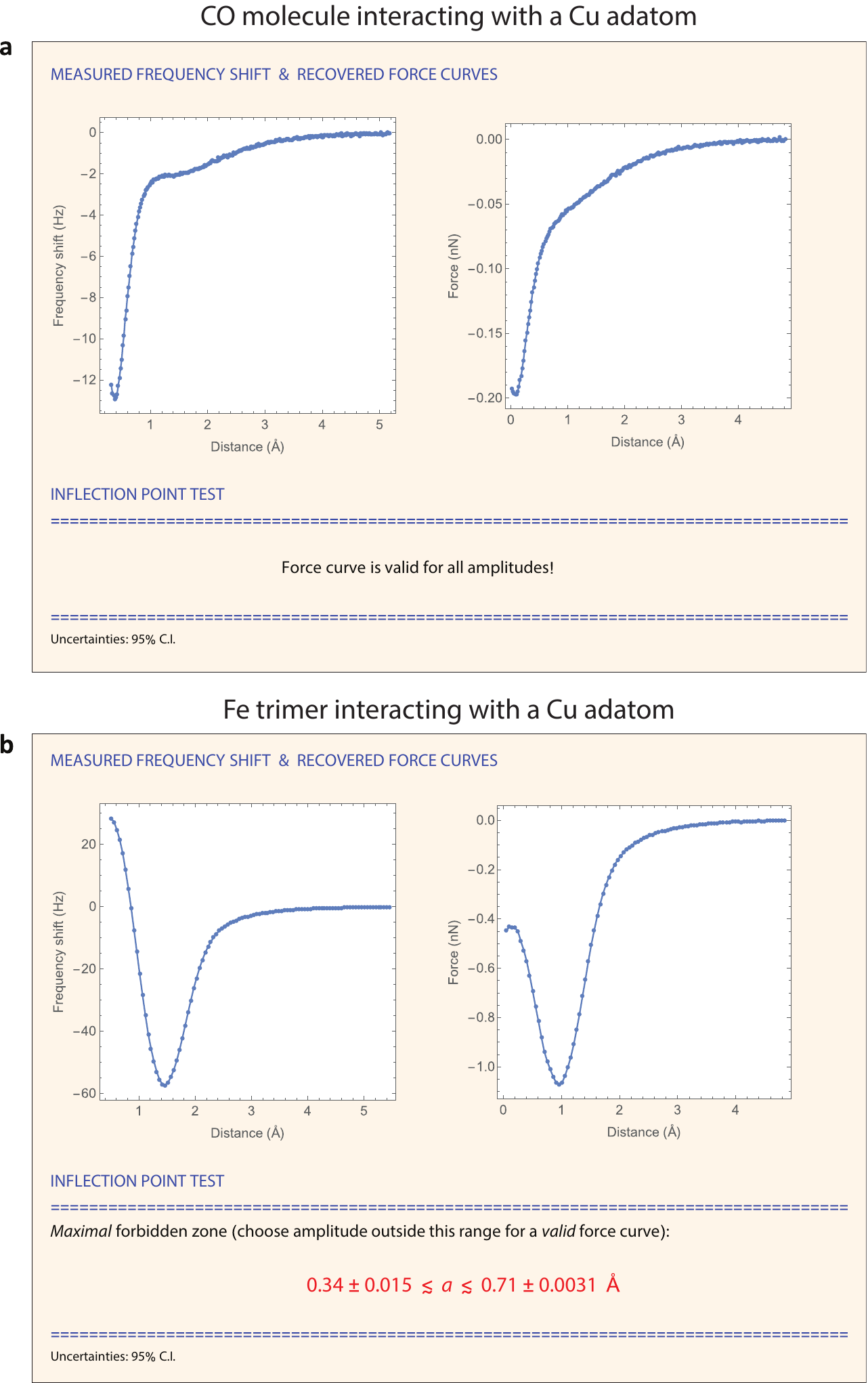}
	\caption{\rm Software output for two interatomic-force measurements. FM-AFM interatomic-force measurements of a Cu adatom on a Cu(111) surface, with \textbf{a}, CO molecule at a tip amplitude of 30 pm, and \textbf{b}, Fe trimer using a tip amplitude of 50 pm. Frequency shift data is from Fig.~\ref{fig:expHuber}.
	}
	\label{fig:software}
\end{figure*}

We first choose the amplitude, $a = 30$\,\,pm, for which the software shows there is one inflection point that exhibits a significant jump in the force: $z_\text{inf} = 0.30 \pm 0.0063\, \, \text{\AA}$ and $z_\text{inf} = 0.30 \pm 0.014 \, \, \text{\AA}$, for the Sader-Jarvis and matrix method recovered forces, respectively. All error bars are 95\% C.I. from the fits to Eq.~(\ref{eq:Taylor}) only. The respective $S$-factors are $S(F) = -0.81 \pm 0.095$ and $-0.77 \pm 0.21$. These values are consistent and satisfy Eq.~(\ref{eq:S}), i.e., $S(F) \gtrsim -1$, establishing that the recovered forces are \textit{valid} for all oscillation amplitudes. The software reports the measured frequency shift data, the recovered force together with an assessment of its validity; see Fig.~\ref{fig:software}a. It informs that the FM-AFM force measurements is valid for any chosen oscillation amplitudes.

This feature is highlighted in Fig.~\ref{fig:expHuber}(a)(middle \& right) which shows the recovered short-range $F(z)$ curves for all five measurement amplitudes, $a = 10$, 30, 50, 75, 100\,\,pm; obtained from the corresponding measured $\Delta f(z)$ curves in Fig.~\ref{fig:expHuber}(a)(left). All recovered force curves coincide to within experimental error; uncertainty in the measurement amplitudes of 5--10\% is not atypical.~\cite{Sader2018}
Coincidence of valid recovered  force curves using different amplitudes is critical in establishing a \textit{robust} force measurement, which clearly occurs here. The datasets reported by the software for all measured oscillation amplitudes are given in Section 3 of Supplementary Material.

This demonstrates the utility of this automation in assessing valid force recovery from the measured frequency shift and obtaining a robust force measurement.

\subsection{Fe trimer/Cu(111) dataset}

Next, we analyze complementary measurements of spectra over the center of a single Fe trimer on Cu(111).~\cite{Huber2020}
Figure~\ref{fig:expHuber}(b)(left) shows the short-range frequency shift curves, $\Delta f (z)$, measured using four different amplitudes.

We choose the $a = 50$\,\,pm frequency measurement, invert it using the Sader-Jarvis and matrix methods, and make an assessment of their validity. In contrast to CO/Cu(111), the Sader-Jarvis recovered force has two inflection points,
\begin{align}
&z_\text{inf} =0.55 \pm 0.0062 \,\, (\text{\AA}) \!: \quad S(F) = -0.93 \pm 0.069,\nonumber\\
&z_\text{inf} =1.4 \pm 0.0062 \,\, (\text{\AA}) \!: \quad S(F) = -4.4 \pm 0.34.
\end{align}
The recovered force using the matrix method gives similar results; see Table S1 in Section 3 of Supplementary Material. Equation~(\ref{eq:S}) shows that the inflection point at $z_\text{inf} \approx 1.4 \,\, \text{\AA}$ can generate ill-posed behavior, since $S(F) \lesssim -1$, with the software reporting a maximal forbidden zone of (Fig.~\ref{fig:software}b)
\begin{gather}
\label{maximalFe}
0.34 \pm 0.015  \lesssim a  \lesssim 0.71 \pm 0.0031\, \, (\text{\AA}).
\end{gather}
The chosen measurement amplitude, $a = 50$\,\,pm, lies within this maximal forbidden zone, indicating ill-posed behavior may be present. A notable difference is observed between recovered force curves from the Sader-Jarvis and matrix methods that exceeds known inversion errors.~\cite{Sader2004,Welker2012b,Dagdeviren2018,SJ2}

An amplitude chosen outside of the forbidden zone in Eq.~(\ref{maximalFe}) can lead to a valid force-distance curve. Indeed, the software reports \textit{valid} force curves for measurements using $a=10$ and 100\,\,pm (see Table S1 in Section 3 of Supplementary Material), which is independently true for force recovery using the Sader-Jarvis and  matrix methods. Moreover, these force curves overlap to within experimental error for all $z$; see Fig.~\ref{fig:expHuber}(b)(middle, right). As noted above, this coincidence of valid force curves using different amplitudes  is vital in establishing that  a \textit{robust} force curve has been achieved. In contrast, the measurement using $a = 30$\,\,pm is again within the forbidden zone (Table S1 of Supplementary Material), and the recovered force curves may not be valid---this measurement must be excluded.~\cite{SJ2}

It is well known that direct discretization, such as used in the matrix method, generally produces spurious solutions in the ill-posed case.~\cite{Phillips1962,Hansen1992,Hansen1998,Vogel2002} For the measurement in Fig.~\ref{fig:expHuber}(b)(right), the matrix method gives similar force-distance curves for all reported amplitudes, even in the forbidden zone. Caution must be exercised when interpreting this observation.
Measurements in Ref.~\onlinecite{Sader2018} and simulations in Fig.~S2 of Supplementary Material show the opposite behavior, with the matrix method producing stronger variations and larger errors with amplitude than the Sader-Jarvis method in the forbidden zone.~\cite{Sader2018} Regularization is required in such cases, for which the Sader-Jarvis method provides only a partial implementation.~\cite{SJ3}
Further work is needed to fully regularize force recovery in FM-AFM; see Ref.~\onlinecite{Sader2018} for details. Therefore, the reporting of force measurements in the forbidden zone using (currently available) methods must be avoided. The inflection point test, its associated algorithm and automated software reported here, provide practitioners with a user-friendly tool to perform valid and robust force measurements.

\section{Conclusions}

In summary, the inflection point test's framework has been further developed and new efficient algorithms formulated that together enable its complete  automation. Software implementing this automation, combined with the Sader-Jarvis method for force recovery, is  provided in the Supplementary Material. This gives all AFM users the ability to easily perform valid and robust force measurements, which was demonstrated on two discrete experimental datasets. While users typically choose a single amplitude for their force measurements, the present study highlights the importance of selecting a range of amplitudes and testing for consistency amongst valid force measurements. Only through this procedure can a robust force measurement be established. Our study suggests that this constitutes good practice for all future measurements.

We conclude by noting that other dynamic AFM modes, e.g., amplitude modulation AFM, are expected to exhibit similar ill-posed behavior. This is because they also blur the underlying force during measurement---due to their inherent use of nonzero oscillation amplitude---and their force recovery involves a deblurring process.~\cite{Sader2018} Inversion formulas identical in form to the Sader-Jarvis method~\cite{Sader2004} for FM-AFM force recovery are used.~\cite{Katan2009,Payam2015} Investigation of these complementary AFM modes represents an interesting avenue for future work.~\cite{Dagdeviren2019}\\

\noindent\textbf{\large{Supplementary Material}}\\

See Supplementary Material for the software, a user guide and fit data for Section~\ref{demonstration}.\\

\noindent\textbf{\large{Acknowledgements}}\\

The author thanks Ferdinand Huber and Franz Giessibl for providing their experimental data and interesting discussions, which motivated this work. Support from the Australian Research Council Centre of Excellence in Exciton Science (CE170100026) and the Australian Research Council Grants Scheme is also gratefully acknowledged. 
Any updates to the software will be available from the author. Data supporting the findings of this study are available from the author upon reasonable request.


%

\end{document}